\documentclass[preprint,aps,prc,showpacs,nofootinbib]{revtex4}
\usepackage{epsfig}
\usepackage{graphicx,amsmath}

\newcommand\ba{\begin{eqnarray}}
\newcommand\ea{\end{eqnarray}}
\newcommand\nn{\nonumber}
\newcommand{\be}{\begin{equation}}
\newcommand{\ee}{\end{equation}}

\begin{document}
\title{Model--independent analysis of polarization effects in elastic
electron--deuteron scattering in presence of two--photon exchange}
\author{G. I. Gakh}
\email{gakh@kipt.kharkov.ua}
\altaffiliation{Permanent address: \it National Science Center "Kharkov Institute of Physics
and Technology", 61108 Akademicheskaya 1, Kharkov, Ukraine}
\affiliation{ DAPNIA/SPhN, CEA/Saclay, 91191 Gif-sur-Yvette Cedex, France}

\author{E. Tomasi-Gustafsson}
\affiliation{ DAPNIA/SPhN, CEA/Saclay, 91191 Gif-sur-Yvette Cedex, France}

\vspace{0.5cm}
\begin{abstract}
The general spin structure of the matrix element, taking into account the
two--photon exchange contribution, for the elastic electron (positron)
--deuteron scattering has been derived using general symmetry properties of
the hadron electromagnetic interaction, such as P--, C-- and T--invariances
as well as lepton helicity conservation in QED at high energy. Taking into account also crossing symmetry, the amplitudes of $e^{\mp}d-$scattering can be
parametrized in terms of fifteen real functions. The expressions for the differential cross
section and for all polarization observables are given in terms of these
functions. We consider the case of an arbitrary polarized
deuteron target and polarized electron beam (both longitudinal
and transverse). The transverse polarization of the electron beam
induces a single--spin asymmetry which is non--zero in presence of
two--photon exchange. It is shown that elastic deuteron electromagnetic form factors can still be extracted in presence of two photon exchange, from the
measurements of the differential cross section and of one polarization
observable (for example, the tensor asymmetry) for electron and
positron deuteron elastic scattering, in the same kinematical conditions.
\end{abstract}

\vspace{0.2cm}
\pacs{25.30.Bf, 13.40.-f, 13.88.+e}
\vspace{0.2cm}
\maketitle
\section{Introduction}
\hspace{0.7cm}

The study of the structure of hadrons and nuclei with electromagnetic probes
is based on the validity of the one--photon exchange mechanism for elastic
and inelastic electron--hadron scattering. This approach is
valid when the possible two--photon exchange (TPE) contribution is small.

Recently it has been possible to apply the polarization method \cite{Re68} for the measurements of the electromagnetic nucleon form factors (FFs) at
high transfer momentum squared $q^2$  \cite{Jo00}. Very precise results were obtained for the ratio of the proton electric and magnetic FFs  which differ from unpolarized cross section measurements 
(Rosenbluth fit)  \cite{AC}. This discrepancy increases when $q^2$ values increase.

In a recent experiment at Jefferson Laboratory \cite{Qa05} a precise
Rosenbluth extraction of the proton FFs, detecting  the recoil
proton, confirms this discrepancy. It was suggested that the
presence of TPE contribution as large as 5\% could solve this problem \cite{Gu03}. Model calculations may give TPE contribution to the cross
section of $ep$ elastic scattering of the order of few percent \cite{Bl03}. A parton model calculation
\cite{Ch04} leads to a quantitative agreement between the standard Rosenbluth
fit and the polarization transfer measurements. Recently the
TPE contribution has been studied for the case of the inelastic
electron--nucleon scattering, $eN\to e\Delta (1232),$ with the aim
of a precision study of the ratios of electric quadrupole $(E2)$
and Coulomb quadrupole $(C2)$ to the magnetic dipole $(M1)$
$\gamma^*N\Delta $ transitions \cite{PCV}. Inelastic intermediate state as $\Delta$ has been calculated in \cite{Ko05,By06}, and it was found to have opposite sign than the proton intermediate state. It has also been argued that elastic and inelastic contributions eventually cancel \cite{By06}. 

An exact calculation has been done in frame of QED and shows that the TPE contribution does not exceed 1\% \cite{Ku06}. A recent calculation of the box diagram in $ep$ elastic scattering \cite{Bo06} also shows that the contribution of this diagram is very small. Higher order radiative corrections, based on the structure function method, when applied to the unpolarized cross section, can bring the results in quantitative agreement \cite{By06}, while the corrections on the polarization ratio are small as well as the TPE contribution. 

Note also that the search of
the deviation from linearity of the Rosenbluth fit to the differential cross
section, using the most recent data on elastic electron--proton scattering,
does not show evidence for such deviation \cite{TG05}. 

A model independent study of the TPE mechanism in the elastic
electron--nucleon scattering and its consequences on the experimental observables, has been carried on  in  \cite{Re04a,Re04b,Re04c}, and in the crossed channels: proton--antiproton annihilation into lepton pair \cite{Ga05}
and annihilation of $e^+e^--$pair into nucleon--antinucleon \cite{Ga06}.

The fact that the TPE mechanism, where the momentum transfer is equally shared between the two virtual photons, can become important when  $q^2$  increases, was already indicated more than thirty years ago \cite{Gu73,Fr73,Bo73}.

Estimates of the TPE contribution to the elastic electron--deuteron
scattering were made in Refs. \cite{Gu73,Fr73} within the framework of the
Glauber theory. It was shown \cite{Gu73} that this contribution decreases
very slowly with increase of the momentum transfer squared
$q^2$ and may dominate in the
cross section at high $q^2$ values. Since the TPE amplitude is essentially
imaginary in this model, the difference between positron and electron
scattering cross sections depends upon the small real part of the TPE
amplitude \cite{Gu73}. Recoil polarization effects may be substantial, in
the region where the one-- and two--photon exchange contributions are
comparable. If the TPE mechanism become sizable, the extraction of the nucleon or deuteron electromagnetic FFs from the experimental data would only be possible after the determination of several (polarization) observables.

It is known that double scattering dominates in collisions of high--energy
hadrons with deuterons at high $q^2$ values, and 
it was predicted that the TPE contribution in the elastic electron--deuteron
scattering contributes for 10 \%  at $|q^2|~\cong ~1.3$  GeV$^2$ \cite{Fr73}. At the
same time the importance of the TPE mechanism was considered in Ref.
\cite{Bo73}.

Using high precision data on the elastic electron--deuteron scattering,
recently obtained at the Jefferson Laboratory, the authors of the paper
\cite{Re99} looked for a possible contribution of the TPE mechanism at
relatively high momentum transfer. While they did not found a definite
evidence for the presence of the TPE contribution in the elastic electron
--deuteron scattering, it was the first attempt to obtain a quantitative
upper limit of a possible TPE contribution using a parametrization of the
TPE term and the existing experimental data. The discrepancy among the set
of data \cite{Al99,Ab99} was therefore attributed to systematic effects.

Tests of the limits of the validity of the one--photon approximation have
been done in the past, using different methods, but no effect has been found
within the accuracy of the performed experiments. At first the TPE contribution was
experimentally observed in the domain of very small energies in
atomic physics \cite{H98} and later the  measurements of the beam asymmetry in the
scattering of transversally polarized electrons on unpolarized protons, performed at MIT (Bates) \cite{We01} and at MAMI (Mainz) \cite{Ma05},   gave small but 
non--zero values for this observable, contrary to what is expected in the
one--photon--exchange (Born) approximation. This asymmetry is related to the
imaginary part of the interference between one-- and two--photon exchange
amplitudes and can be connected only indirectly with the real part of this
interference which contribute to the differential cross section
or to the T--even polarization observables of the elastic electron--hadron scattering.

The recoil--deuteron vector polarization for the elastic scattering of
electrons from an unpolarized deuteron target, which vanishes in the Born
approximation, was measured in an experiment which aimed to test time
--reversal invariance in electromagnetic interaction at high momentum
transfer. The value obtained for the vector polarization was $|P|=$0.075 $
\pm $0.088 \cite{Pr68}, and the precision of the experiment did not allow
to find evidence for the TPE contribution.

In this paper we analyze, in a model--independent way, the influence of the
TPE contribution on the differential cross section and various polarization
observables in the elastic electron (positron)--deuteron scattering
\be e^{\mp}+d\to e^{\mp}+d. \label{Eq:eq1} \ee

Our approach is similar to the one used earlier for the analysis of the TPE
mechanism in the elastic electron--nucleon scattering \cite{Re04a}. The
situation with elastic $e^{\mp}d-$ scattering is more complicated in
comparison with the elastic $eN-$ scattering since the deuteron has spin one.
In this case the spin structure of the matrix elements of the reactions
(\ref{Eq:eq1}) are completely determined by fifteen real functions: six
complex amplitudes (in comparison with three complex amplitudes for the case
of elastic $eN-$scattering) depending on two variables and three deuteron
electromagnetic FFs (two nucleon FFs for elastic $eN-$scattering) which are real functions of one variable, $Q^2$.

The purpose of this paper is to derive general expressions for the
differential cross section and various polarization observables in elastic
electron (positron) --deuteron scattering and to suggest model independent
methods to extract deuteron electromagnetic FFs also in presence of the TPE
contribution, without any underlying assumptions.

Note that the presence of the TPE contribution in $e^{\mp}+d\to e^{\mp}+d$
results in nonlocal spin structure of the matrix element. The
standard analysis of the polarization effects, which is known for the
one--photon--exchange mechanism, does not apply anymore. The extraction of form factors can be done after a more complicated derivation involving additional polarization observables.

\section{Matrix element and symmetry relations}
\hspace{0.7cm}

The starting point of our analysis is the following general parametrization
of the spin structure of the matrix element for elastic electron--deuteron
scattering, which can be obtained from the non--spin--flip part of the
amplitude of the elastic nucleon--deuteron scattering \cite{CRMS}
\be\label{Eq:eq2}
{\it M} =\frac{e^2}{Q^2}l_{\mu }J_{\mu }. \ee
The leptonic and hadronic currents have the form
\be\label{Eq:eq3}
l_{\mu }=\bar u(k_2)\gamma_{\mu }u(k_1), \ \ee
$$J_{\mu }=(p_1+p_2)_{\mu }\Bigg [-G_{1}(s, Q^2)U_1\cdot U_2^*+
\frac{1}{M^2}G_{3}(s, Q^2)(U_1\cdot qU_2^*\cdot q
-\frac{q^2}{2}U_1\cdot U_2^*)\Bigg ]+ $$
$$+G_2(s,Q^2)(U_{1\mu }U_2^*\cdot q-U_{2\mu }^*U_1\cdot q)+
\frac{1}{M^2}(p_1+p_2)_{\mu }\Bigg [G_4(s,Q^2)U_1\cdot kU_2^*\cdot k+ $$
\be\label{Eq:eq4}
+G_5(s,Q^2)(U_1\cdot qU_2^*\cdot k-U_1\cdot kU_2^*\cdot q)\Bigg ]+
G_6(s,Q^2)(U_{1\mu }U_2^*\cdot k+U_{2\mu }^*U_1\cdot k), \ee
where $k_1~(k_2)$ and $p_1~(p_2)$ are the four--momenta of the initial
(scattered) electron and initial (scattered) deuteron, respectively;
$k=k_1+k_2,$ $q=k_1-k_2=p_2-p_1,$ $Q^2=-q^2,$ $M$ is the deuteron mass,
$U_{1\mu}(U_{2\mu})$ is the initial (final) deuteron polarization
four--vector.

The six complex amplitudes, $G_{i}(s,Q^2), \ i=1-6$, which are generally
functions of two independent kinematical variables, $Q^2$ and $s=(k_1+p_1)^2$
($s$ is the square of the total energy of the colliding particles), fully
describe the spin structure of the matrix element for the elastic electron
--deuteron scattering - for any number of exchanged virtual photons.

In the Born (one--photon--exchange) approximation these amplitudes reduce to
three:
\be\label{Eq:eq5}
G_{1}^{Born}(s,Q^2)=G_{1}(Q^2), \  G_{2}^{Born}(s,Q^2)=G_{2}(Q^2),
\ G_3^{Born}(s,Q^2)=G_3(Q^2), \ \ee
$$G_{i}^{Born}(s,Q^2)=0, \ \ i=4,5,6, $$
where $G_i(Q^2)$, $(i=1, 2, 3),$ are the deuteron
electromagnetic FFs depending only on the virtual photon four--momentum
squared. Due to the current hermiticity, FFs $G_i(Q^2)$ are real functions
in the region of the space--like momentum transfer. The same FFs describe
also the one--photon--exchange mechanism for  elastic positron--deuteron
scattering.

These FFs are related to the standard deuteron FFs: $G_C(Q^2)$ (the charge
monopole), $G_M(Q^2)$ (the magnetic dipole) and $G_Q(Q^2)$ (the charge
quadrupole). These relations are
\be\label{Eq:eq6}
G_M(Q^2)=-G_2(Q^2), \ G_Q(Q^2)=G_1(Q^2)+G_2(Q^2)+2G_3(Q^2), \ \ \ee
$$G_C(Q^2)=\frac{2}{3}\tau [G_2(Q^2)-G_3(Q^2)]+(1+\frac{2}{3}\tau )
G_1(Q^2), $$
with
$$\tau =\frac{Q^2}{4M^2}. $$
The standard FFs have the following normalizations:
\be\label{Eq:eq7} G_C(0)=1\ , \ \  G_M(0)=(M/m)\mu_d\ ,
\ \ G_Q(0)=M^2Q_d\ , \ee
where $m$ is the nucleon mass, $\mu_d= 0.857(Q_d=0.2859)$ is deuteron magnetic (quadrupole)
moment. The numerical values are taken from \cite{MT,ERC}.

The spin structure of the matrix element for the elastic electron--deuteron
scattering can be established in analogy with the elastic nucleon--deuteron
scattering \cite{CRMS}, using the general properties of the electron--hadron
interaction, such as the Lorentz invariance and P--invariance. Taking into
account the identity of the initial and final states and the T--invariance
of the electromagnetic interaction, the reactions $e^{\mp}+d\to e^{\mp}+d$,
where a spin $1/2$ particle is scattered by a spin $1$ particle, are
described by twelve independent complex amplitudes. So, the model--independent
parametrization of the corresponding matrix element can be done (in many
different but equivalent forms) in terms of twelve invariant complex
amplitudes, $G_{i}(s,Q^2)$, $i=1-12$.

At high energies we can neglect the contributions which are proportional to
the electron mass. In this limit, any Feynman diagram in QED is invariant
under the chirality operation $u(p)\to \gamma_5u(p)$. This invariance implies
that invariant structures in the matrix element which change their sign under
this transformation can be neglected since they are proportional to the
electron mass. So, the structures as $\bar u(k_2)u(k_1)$ and
$\bar u(k_2)\gamma_{\mu }\gamma_{\nu }u(k_1)$ can be neglected. As a result,
we can reduce the number of invariant amplitudes for elastic
electron--deuteron scattering from twelve amplitudes to six ones.

Let us stress that in the general case

-- The amplitudes $G_{i}(s,Q^2)$, $i=1-6$, are the complex functions of two
independent variables, $Q^2$ and $s$.

-- The connection of these amplitudes with the deuteron electromagnetic
FFs is non--trivial since the amplitudes are related to the amplitudes of
the virtual Compton scattering process which are presently unknown.

-- The set of the amplitudes $G_i^{(-)}(s, Q^2)$ for the reaction
$e^-+d\to e^-+d$ is different from the corresponding set of the amplitudes
$G_i^{(+)}(s, Q^2)$ describing the charge conjugated reaction $e^++d\to e^++d$. This means that the properties of the
positron--deuteron scattering cannot be derived from the
$G_i^{(-)}(s, Q^2)$ amplitudes. However, prescriptions based on C--invariance help to derive  expressions which rely real FFs, which are functions of $Q^2$, to experimental observables. The strategy for their determination in presence of TPE will be detailed below.

Let us introduce another set of variables: $\epsilon $ and $Q^2$, which is
equivalent to $s$ and $Q^2$ (in Lab system):
$$\epsilon^{-1}=1+2(1+\tau )\tan^2\frac{\theta}{2}, $$
where $\theta $ is the electron scattering angle in the Lab system. These
variables $\epsilon $ (the degree of the linear polarization of the virtual
photon) and $Q^2$ are well suited for the description of the
electron--hadron elastic scattering in the one--photon--exchange
approximation, since only the $Q^2-$dependence of FFs has a dynamical origin,
whereas the linear $\epsilon $--dependence of the differential cross section
is a consequence of the one--photon mechanism. The variables $s$ and $Q^2$
are more convenient for the annihilation channel and for the analysis of the
consequences of the crossing symmetry.

To separate the effects caused by the Born (one--photon exchange) and TPE
contributions, let us single out the dominant contribution and define the
following decompositions of the amplitudes (taking into account the
C--invariance of the electromagnetic interaction of hadrons)
\be\label{Eq:eq8}
G_i^{(\mp )}(Q^2, \epsilon )=\mp G_i(Q^2)+\Delta G_i(Q^2, \epsilon ),
\ \ i=1, 2, 3, \ee \be\label{Eq:eq9}
G_i^{(-)}(Q^2, \epsilon )=G_i^{(+)}(Q^2, \epsilon )=G_i(Q^2, \epsilon ),
\ \ i=4, 5, 6, \ee
where $\Delta G_{1,2,3}$ and $G_{4,5,6}$ describe the TPE contribution only.

The order of magnitude of these quantities is $\Delta G_{i}(Q^2, \epsilon )$,
$(i=1,2,3),$ and $G_i(Q^2, \epsilon )$, $(i=4,5,6),$ $ \sim\alpha $,
and $G_{i}(Q^2)$, $ (i=1,2,3),$ $\sim \alpha^0.$ Since the terms
$\Delta G_{i}$, $(i=1,2,3), $ and $G_i $, $(i=4,5,6),$ are small in
comparison with the dominant ones, we neglect in following by the bilinear
combinations of these small terms.

Therefore the reactions $e^{\mp}+d\to e^{\mp}+d$ are described by fifteen
different real functions:

- three real FFs $G_i(Q^2) \ (i=1-3)$, which are functions of one variable
only. This holds in the space--like region since in the time--like region
these FFs became complex functions due to the strong interaction in the
final state as in the case of $e^++e^-\to$ $\rho^+ +\rho^-$, $\bar d +d$
reactions \cite{Ga06a}.

- six functions: $\Delta G_{1,2,3}(Q^2, \epsilon )$ and $G_{4,5,6}(Q^2,
\epsilon ),$  which are, in the general case, complex functions of two variables,
$Q^2$ and $\epsilon $.

We will use the following notations
\ba  G_M^{(\mp )}(Q^2, \epsilon )&=&-G_2^{(\mp )}(Q^2, \epsilon ), \
G_Q^{(\mp )}(Q^2, \epsilon )=G_1^{(\mp )}(Q^2, \epsilon )+G_2^{(\mp )}(Q^2,
\epsilon )+2G_3^{(\mp )}(Q^2, \epsilon ), \ \ \nonumber\\
G_C^{(\mp )}(Q^2, \epsilon )&=&\frac{2}{3}\tau [G_2^{(\mp )}(Q^2, \epsilon )
-G_3^{(\mp )}(Q^2, \epsilon )]+(1+\frac{2}{3}\tau )G_1^{(\mp )}
(Q^2, \epsilon ). \label{Eq:eq10} \ea
So, the quantities $G_i^{(\mp )}(Q^2, \epsilon ),$ $i=M, Q, C$ can be
considered as generalized magnetic, quadrupole and charge FFs.

We can separate the Born and TPE contributions, in these generalized FFs, in the following way
\be\label{Eq:eq11}
G_i^{(\mp )}(Q^2, \epsilon )=\mp G_i(Q^2)+\Delta G_i(Q^2, \epsilon ),
\ \ i=M, Q, C, \ee
where $\Delta G_i(Q^2, \epsilon )$ contain the TPE contribution.


\section{General analysis}

\hspace{0.7cm}

In the Laboratory (Lab) system, including the contribution of the TPE
mechanism, the unpolarized differential cross section for elastic
$e^{\mp}d-$ scattering can be written as
\be\label{Eq:eq12}
\frac{d\sigma}{d\Omega}=\frac{\alpha^2}{4M^2Q^4}\frac{E'}{E}\left
(1+2\frac{E}{M}sin^2\frac{\theta}{2}\right  )^{-1}L_{\mu\nu }H_{\mu\nu }, \
\ee
with
$$L_{\mu\nu }=l_{\mu }l_{\nu}^*, \ \  H_{\mu\nu }=J_{\mu }J_{\nu}^*, $$
where $E(E')$ is the energy of the initial (scattered) electron or positron.
Here and below we neglect the electron mass where it is possible.

The leptonic tensor, for the case of polarized electron or positron beam,
has the form
\be\label{Eq:eq13}
L_{\mu\nu }=q^2g_{\mu\nu }+2(k_{1\mu }k_{2\nu }+k_{1\nu }k_{2\mu })+
2im_e<\mu\nu qs_e>, \ \ee
where $<\mu\nu ab>=\varepsilon_{\mu\nu\rho\sigma }a_{\rho}b_{\sigma}$,
$m_e$ is the electron mass and $s_{e\mu}$ is the polarization four--vector
of the initial electron or positron.

Since we consider only the case of the polarized target, the hadronic tensor
can be expanded according to the polarization state of the initial deuteron
as follows:
\be\label{Eq:eq14}
H_{\mu\nu }=H_{\mu\nu }(0)+H_{\mu\nu }(s)+H_{\mu\nu }(Q), \ee
where the tensor $H_{\mu\nu }(0)$ corresponds to the unpolarized target,
the tensor $H_{\mu\nu }(s)(H_{\mu\nu }(Q))$ describes the case when deuteron
target has vector (tensor) polarization.

The spin--density matrix of the initial (polarized) and recoil (unpolarized)
deuterons can be written as
\be\label{Eq:eq15}
\rho_{i\mu\nu}=U_{1\mu}U_{1\nu}^*=-\frac{1}{3}(g_{\mu\nu}-
\frac{p_{1\mu}p_{1\nu}}{M^2})
+\frac{i}{2M}<\mu\nu sp_1>+Q_{\mu\nu}, \ee
$$\rho_{f\mu\nu}=U_{2\mu}U_{2\nu}^*=-(g_{\mu\nu}-
\frac{p_{2\mu}p_{2\nu}}{M^2}), $$
where $s_{\mu}$ is the polarization four--vector describing the vector
polarization of the target ($p_1\cdot s=0, s^2=-1$) and $Q_{\mu\nu}$ is the
tensor describing the tensor (quadrupole) polarization of the target
($Q_{\mu\nu}=Q_{\nu\mu}$, $Q_{\mu\mu}=0$, $p_{1\mu}Q_{\mu\nu}=0$). In Lab
system all time components of the tensor $Q_{\mu\nu}$ are zero and the
tensor polarization of the target is described by five independent space
components ($Q_{ij}=Q_{ji}, Q_{ii}=0, i,j=x,y,z$).

In the hadronic current $J_{\mu}$, the presence of the TPE contribution
leads to the terms which contain the momenta from the leptonic vertex. The
general structure of the tensor $H_{\mu\nu}(0)$ becomes more complicated:
four structure functions are present instead of the two standard structure
functions $A(Q^2)$ and $B(Q^2)$. The general structure of this tensor can be
written as
\be\label{Eq:eq16}
H_{\mu\nu }(0)=H_1\tilde g_{\mu\nu }+H_2p_{\mu }p_{\nu }+H_3(k_{\mu }p_{\nu }
+k_{\nu }p_{\mu })+iH_4(k_{\mu }p_{\nu }-k_{\nu }p_{\mu }), \ee
where $\tilde g_{\mu\nu }=g_{\mu\nu }-q_{\mu }q_{\nu}/q^2,$ $p=p_1+p_2.$ One
can get the following expressions for these structure functions when the
hadronic current is given by Eq. (\ref{Eq:eq4}):
\ba H_1&=&\frac{2}{3}q^2\left [(1+\tau )|G_{2}|^2-aReG_{2}G_{6}^*\right  ],
\nn \\ H_2&=&(1+2\tau )|G_{1}|^2-\frac{8}{3}\tau Re(G_{4}+2aG_5)G_{1}^* +
\nn \\ &&+\frac{2}{3}(2\tau -1)Re(\tau G_{1}+2\tau G_3-a^2G_4-4\tau aG_5-
aG_6)G_{1}^* +\nn \\ &&+\frac{2}{3}\tau Re\left [(1+2\tau )(2G_1+G_2-4aG_5)
-2a(aG_4+G_6)\right  ]G_2^*-\frac{4}{3}\tau aRe(aG_4+G_6)G_3^*+ \nn \\
&&+4\tau^2\left  [|G_{3}|^2+\frac{2}{3}Re(2G_4+G_2-2aG_5)G_3^*\right  ],\nn
\\ H_3&=&\frac{2}{3}Re(\tau G_{2}-G_1+2\tau G_3)G_{6}^*+\frac{4}{3}\tau
Re\left [aG_4+2(1+\tau )G_5\right  ]G_2^*,\nn \\ H_4&=&\frac{2}{3}
Im(\tau G_{2}+G_1-2\tau G_3)G_{6}^*-\frac{4}{3}\tau Im\left [aG_4+
2(1+\tau )G_5\right  ]G_2^*, \label{Eq:eq17} \ea
where $a=k\cdot p_1/M^2$. One can see that the structure functions $H_3$ and
$H_4$ are completely determined by the TPE contribution. We recover the
standard tensor structure for $H_{\mu\nu}(0)$ tensor, if the TPE
contribution is absent.

Let us consider the part of the hadronic tensor that corresponds to the
vector--polarized deuteron target. It can be represented as the sum of a
symmetrical and antisymmetrical tensors (with respect to the indexes $\mu$
and $\nu $ ):
\be\label{Eq:eq18}
H_{\mu\nu}(s)=iA_{\mu\nu}(s)+S_{\mu\nu}(s), \ee
where the antisymmetrical tensor $A_{\mu\nu}(s)$ can be written as (we
neglect the terms proportional to $q_{\mu}$ or $q_{\nu}$ since the leptonic
tensor is conserved, so these terms do not contribute to the observables)
\ba
A_{\mu\nu}(s)&=&A_1<\mu\nu sp_1>+A_2(p_{1\mu}p_{2\nu}-p_{1\nu}p_{2\mu})+
A_3(p_{\mu}k_{\nu}-p_{\nu}k_{\mu})+A_4(A_{\mu}k_{\nu}-A_{\nu}k_{\mu})+
\nn \\  && +A_5(p_{\mu}A_{\nu}-p_{\nu}A_{\mu})+A_6(p_{2\mu}A_{\nu}-
p_{2\nu}A_{\mu})+A_7(p_{\mu}B_{\nu}-p_{\nu}B_{\mu})+
\nn \\ && +A_8(p_{2\mu}B_{\nu}-p_{2\nu}B_{\mu}),
\label{Eq:eq19} \ea
with $A_{\mu}=<\mu p_1p_2s>$, $B_{\mu}=<\mu p_1ks>$, and the structure
functions $A_{1-8}$ can be written as
\ba
A_1&=&2M\tau \left [(1+\tau )|G_{2}|^2-aReG_{2}G_{6}^*\right  ], \nn \\
A_2&=&\frac{b}{2M^3}Re\left [(2\tau G_5+aG_4)G_2^*-G_1G_6^*\right  ],
\nn \\  A_3&=&-\frac{b}{2M^3}ReG_2G_4^*, \ \  A_4=\frac{1}{2M}ReG_2G_6^*,
\nn \\ A_5&=&\frac{1}{2M}Re\left [-2(1+\tau )G_1G_2^*+2a\tau G_2G_5^*+
aG_1G_6^*\right  ],  \nn \\ A_6&=&-\frac{1}{2M}(2\tau |G_{2}|^2-
aReG_{2}G_{6}^*),  \nn \\ A_7&=&\frac{\tau}{M}Re[aG_4+2(1+\tau )G_5]G_{2}^*,
\nn \\ A_8&=&-\frac{a}{2M}ReG_{2}G_{6}^*, \ \  b=<skp_1p_2>.
\label{Eq:eq20} \ea
The symmetrical tensor $S_{\mu\nu}(s)$ can be represented as (neglecting
again the terms proportional to $q_{\mu}$ or $q_{\nu}$)
\ba
S_{\mu\nu}(s)&=&B_1g_{\mu\nu}+B_2p_{\mu}p_{\nu}+B_3p_{2\mu}p_{2\nu}+
B_4(p_{\mu}p_{2\nu}+p_{\nu}p_{2\mu})+\nn \\ &&+B_5(p_{\mu}k_{\nu}+
p_{\nu}k_{\mu})+B_6(A_{\mu}k_{\nu}+A_{\nu}k_{\mu})+B_7(A_{\mu}p_{\nu}+
A_{\nu}p_{\mu})+ \nn \\ &&+B_8(A_{\mu}p_{2\nu}+A_{\nu}p_{2\mu})+
B_9(p_{\mu}B_{\nu}+p_{\nu}B_{\mu})+B_{10}(p_{2\mu}B_{\nu}+p_{2\nu}B_{\mu}).
\label{Eq:eq21} \ea
The structure functions $B_{1-10}$ can be written as
\ba
B_1&=&-\frac{b}{M}ImG_{2}G_{6}^*, \ \  B_2=\frac{b}{M^3}ImG_{1}[aG_4+
2(1+\tau )G_5]^*,~B_3=\frac{b}{M^3}ImG_{2}G_{6}^*, \nn \\
B_4&=&\frac{b}{2M^3}Im\left [G_1G_6^*-2\tau G_5G_2^*-aG_4G_2^*\right  ],
\ \ B_5=\frac{b}{2M^3}ImG_{4}G_{2}^*, \ \ B_6=\frac{1}{2M}ImG_{2}G_{6}^*,
\nn \\  B_7&=&-\frac{a}{2M}ImG_{1}G_{6}^*+\frac{1}{M}Im[(1+\tau )G_1-
a\tau G_5]G_2^*, \ \, B_8=\frac{a}{2M}ImG_{6}G_{2}^*, \nn \\
B_9&=&-\frac{\tau}{M}Im[aG_4+2(1+\tau )G_5]G_2^*, \ \
B_{10}=\tau\frac{a}{M}ImG_{6}G_{2}^*.
\label{Eq:eq22} \ea
Let us note that symmetric tensor is completely determined by the TPE terms
for the case of the space--like region of the momentum transfer squared
(where all deuteron electromagnetic FFs are real functions). So, this tensor
vanishes in the Born approximation. As it is determined by the product
$Im(G_C+\tau /3G_Q)G_M^*$, it may be non--zero in  time--like region, where
deuteron FFs are complex functions.

Let us consider the part of the hadronic tensor that corresponds to a
tensor--polarized deuteron target. It can be also written as the sum of a
symmetrical and an antisymmetrical tensors:
\be\label{Eq:eq23}
H_{\mu\nu}(Q)=S_{\mu\nu}(Q)+iA_{\mu\nu}(Q). \ee
The symmetrical tensor $S_{\mu\nu}(Q)$ can be written as (we neglect
here the terms proportional to $q_{\mu}$ or $q_{\nu}$ )
\ba
S_{\mu\nu}(Q)&=&R_1Q_{\mu\nu}+R_2(g_{\mu\nu}-\frac{p_{2\mu}p_{2\nu}}{M^2})+
R_3p_{\mu}p_{\nu}+R_4(p_{\mu}k_{\nu}+p_{\nu}k_{\mu})+ \nn \\
&&+R_5(p_{\mu}p_{2\nu}+p_{\nu}p_{2\mu})+R_6(p_{\mu}Q_{1\nu}+p_{\nu}Q_{1\mu})+
R_7(k_{\mu}Q_{1\nu}+k_{\nu}Q_{1\mu})+ \nn \\ &&+R_8(Q_{1\mu}p_{2\nu}+
Q_{1\nu}p_{2\mu})+R_9(p_{\mu}Q_{2\nu}+p_{\nu}Q_{2\mu})+R_{10}(p_{2\mu}
Q_{2\nu}+p_{2\nu}Q_{2\mu}),
\label{Eq:eq24}\ea
where $Q_{1\mu}=Q_{\mu\nu}q_{\nu}$, $Q_{2\mu}=Q_{\mu\nu}k_{\nu}$, and the
structure functions $R_{1-10}$ are
\ba
R_1&=&4M^2\tau \left [(1+\tau )|G_{2}|^2-aReG_{2}G_{6}^*\right  ], \ \
R_2=-Q_1|G_{2}|^2+2Q_{12}ReG_{2}G_{6}^*,\nn \\ R_3&=&\frac{Q_1}{M^2}Re
(G_1+2G_3-2aG_5)G_1^*+2\frac{Q_{11}}{M^2}Re(G_1-2\tau G_3)G_4^*-
2\frac{Q_{12}}{M^2}Re(aG_1G_4^*+ \nn \\ &&+2\tau G_1G_5^*+4\tau G_3G_5^*),
\ \ R_4=\frac{1}{M^2}Re(Q_1G_5+Q_{12}G_4)G_2^*, \nn \\ R_5&=&\frac{1}{M^2}
Re(Q_1G_2-Q_{12}G_6)G_1^*-\frac{1}{M^2}Re[aQ_{12}G_4+(aQ_1+2\tau
Q_{12})G_5]G_2^*, \nn \\R_6&=&Re(2\tau G_2-aG_6)G_1^*+2\tau Re(2G_3-
aG_5)G_2^*, \ \ R_7=ReG_{2}G_{6}^*, \nn \\ R_8&=&2\tau |G_{2}|^2-
aReG_{2}G_{6}^*,  \ \ R_9=2Re(G_1-2\tau G_3)G_6^*-2\tau Re[aG_4+
2\tau (1+\tau )G_5]G_2^*, \nn \\ R_{10}&=&-2\tau ReG_{2}G_{6}^*,
\label{Eq:eq25} \ea
with $Q_1=Q_{\mu\nu}q_{\mu}q_{\nu}$, $Q_{12}=Q_{\mu\nu}q_{\mu}k_{\nu}$, and
$Q_{11}=Q_{\mu\nu}k_{\mu}k_{\nu}$. The antisymmetrical tensor $A_{\mu\nu}(Q)$
has the form (also neglecting the terms proportional to $q_{\mu}$ or $q_{\nu}$ )
\ba
A_{\mu\nu}(Q)&=&W_1(p_{\mu}k_{\nu}-p_{\nu}k_{\mu})+W_2(p_{\mu}p_{2\nu}-
p_{\nu}p_{2\mu})+W_3(p_{\mu}Q_{1\nu}-p_{\nu}Q_{1\mu})+\nn \\ &&
+W_4(k_{\mu}Q_{1\nu}-k_{\nu}Q_{1\mu})+W_5(p_{2\mu}Q_{1\nu}-p_{2\nu}Q_{1\mu})+
W_6(p_{\mu}Q_{2\nu}-p_{\nu}Q_{2\mu})+ \nn \\ && +W_7(p_{2\mu}Q_{2\nu}-
p_{2\nu}Q_{2\mu}), \label{Eq:eq26} \ea
where the structure functions $W_{1-7}$ are
\ba
&&W_1=\frac{1}{M^2}Im(Q_1G_5+Q_{12}G_4)G_2^*,\nn \\
&&W_2=-\frac{1}{M^2}ImG_1(Q_1G_2+Q_{12}G_6)^*+\frac{1}{M^2}ImG_2[aQ_{12}G_4+
(aQ_1+2\tau Q_{12})G_5]^*, \nn \\ &&W_3=ImG_1(2\tau G_2-aG_6)^*+
2\tau Im(2G_3-aG_5)G_2^*,\nn \\ &&W_4=ImG_2G_6^*, \ \  W_5=-aImG_2G_6^*,
\nn \\ &&W_6=2Im(G_1-2\tau G_3)G_6^*-2\tau Im[aG_4+2(1+\tau )G_5]G_2^*,
\nn \\ &&W_7=2\tau ImG_2G_6^*.
\label{Eq:eq27} \ea
For simplicity, we omitted in the hadronic structure functions the upper indexes $(\mp )$ referring to
electron-- or positron--scattering. The expressions for all hadronic structure functions hold in 
both cases and the expression for the amplitudes should be understood as:
$G_i= G_i^{(\mp )}(Q^2, \epsilon )$.


\section{T--even polarization observables}

\hspace{0.7cm}


Let us specify the coordinate frame in the Lab system: the $z$ axis is
directed along the momentum transfer ${\vec q}$ and the momenta of the
initial and scattered electrons lie in the $xz$ plane. The $y$ axis is
directed along the direction of the vector ${\vec q}\times {\vec k}_1$.

The following general formula holds for the differential cross section of
the elastic scattering of an unpolarized electron (positron) beam by an
unpolarized deuteron target (taking into account the TPE contribution at
the level of its interference with the Born term):
\ba
\frac{d\sigma_{un}^{(\mp)}}{d\Omega } &=& \sigma_0N^{(\mp)}(Q^2, \epsilon ),
~N^{(\mp)}(Q^2, \epsilon )=A^{(\mp)}(Q^2, \epsilon )+
B^{(\mp)}(Q^2, \epsilon )\tan^2\frac{\theta}{2},\nn \\
\sigma_0&=&\frac{\alpha^2\cos^2\frac{\theta}{2}}{4E^2\sin^4\frac{\theta}
{2}}\left (1+2\frac{E}{M}sin^2\frac{\theta}{2}\right  )^{-1}.
\label{Eq:eqsigma} \ea
The functions $A^{(\mp)}(Q^2, \epsilon )$ and $B^{(\mp)}(Q^2, \epsilon )$
contain the TPE contribution and they have the following form (the signs
$(\mp )$ correspond to the $e^{(\mp)}d-$ scattering)
$$A^{(\mp)}(Q^2, \epsilon )=A(Q^2)\mp \Delta A(Q^2, \epsilon ),
\ \ B^{(\mp)}(Q^2, \epsilon )=B(Q^2)\mp \Delta B(Q^2, \epsilon ), $$
where the structure functions $A(Q^2)$ and $B(Q^2)$ are the standard real
functions of a single variable $Q^2$ describing the elastic $ed-$scattering
in the Born approximation. They are quadratic combinations of the deuteron
electromagnetic FFs
$$A(Q^2)=G_C^2(Q^2)+\frac{2}{3}\tau G_M^2(Q^2)+\frac{8}{9}\tau^2 G_Q^2(Q^2),
\ \ B(Q^2)=\frac{4}{3}\tau (1+\tau )G_M^2(Q^2). $$
The additional terms $\Delta A(Q^2, \epsilon )$ and $\Delta B(Q^2,
\epsilon )$ are due to the TPE contribution and they can be written as
\ba
&&\Delta A(Q^2, \epsilon )=2G_C(Q^2)Re\Delta G_C(Q^2, \epsilon )+
\frac{4}{3}\tau G_M(Q^2)Re\Delta G_M(Q^2, \epsilon )+\nn \\
&& +\frac{16}{9}\tau^2 G_Q(Q^2)Re\Delta G_Q(Q^2, \epsilon )+ \nn \\
&& +\frac{8}{3}\tau\left\{\frac{2}{3}\tau G_Q(Q^2)-G_C(Q^2)+
c^2\left[(1-\tau )G_C(Q^2)+2\tau G_M(Q^2)-\frac{2}{3}\tau^2G_Q(Q^2)\right ]
\right\}\nn \\ && ReG_4(Q^2, \epsilon )+\frac{16}{3}c\tau \sqrt{\tau (1+\tau )}
(G_M(Q^2)-G_C(Q^2)-\frac{4}{3}\tau G_Q(Q^2))ReG_5(Q^2, \epsilon )+\nn
\\ && +\frac{4}{3}c\sqrt{\frac{\tau }{1+\tau }}\left [(1-\tau )G_C(Q^2)+
2\tau G_M(Q^2)-\frac{2}{3}\tau (1+2\tau )G_Q(Q^2)\right ]ReG_6(Q^2, \epsilon ),
 \nn \\ &&\Delta B(Q^2, \epsilon )=\frac{8}{3}\tau G_M(Q^2)
\left [(1+\tau )Re\Delta G_M(Q^2, \epsilon )+
c\sqrt{\tau (1+\tau )}ReG_6(Q^2, \epsilon )\right ],
\label{Eq:eq29} \ea
where
$$c=\sqrt{\frac{1+\epsilon }{1-\epsilon }}=
\sqrt{1+\frac{\cot^2\frac{\theta}{2}}{1+\tau}}. $$
Note that these formulas were obtained neglecting the terms of the order of
$\alpha ^2$ compared to the dominant (Born approximation) terms. In the Born
approximation these expressions reduce to the well known result for the
differential cross section of elastic $ed-$ scattering (see, for example,
\cite{GG02}).

The structure function $B^{(\mp )}(Q^2, \epsilon )$, which is determined in
the Born approximation by the magnetic FF only, acquires two additional terms
proportional to $Re\Delta G_M(Q^2, \epsilon )$ and $ReG_6(Q^2, \epsilon )$.

The real parts of all six complex TPE amplitudes contribute to the structure
function $A^{(\mp )}(Q^2, \epsilon )$, which is determined in the Born
approximation by all three deuteron FFs.

One can see that the sum of the differential cross sections for the
$e^{\mp}d-$scatterings has precisely the Rosenbluth form, in terms of the
standard deuteron electromagnetic FFs, since the TPE contribution is
canceled out 
\be\label{Eq:eq30}
\Sigma=\frac{1}{2}\left(\frac{d\sigma_{un}^{(-)}}{d\Omega }+
\frac{d\sigma_{un}^{(+)}}{d\Omega }\right )=\sigma_0
\left[A(Q^2)+B(Q^2)\tan^2\frac{\theta}{2}\right  ]. \ee
This quantity allows to separate the magnetic FF $G_M(Q^2)$ and the following
combination of the charge and quadrupole FFs: $G(Q^2)=G_C^2(Q^2)+
\frac{8}{9}\tau^2 G_Q^2(Q^2),$ in presence of the TPE contribution. It is a
model independent statement, taking into account the
interference of the one-- and two--photon exchange amplitudes.

On the contrary, the difference of the differential cross sections for the
$e^{\mp}d-$scatterings is completely determined by the interference of the
one-- and two--photon exchange amplitudes and it can be written as
\be\label{Eq:eq31}
\frac{1}{2}\left (\frac{d\sigma_{un}^{(+)}}{d\Omega }-
\frac{d\sigma_{un}^{(-)}}{d\Omega }\right  )=\sigma_0\left [\Delta A(Q^2,
\epsilon )+\Delta B(Q^2, \epsilon )\tan^2\frac{\theta}{2}\right ]. \ee
This quantity contains information about the size of the TPE term and its
dependence on the variables $Q^2$ and $\epsilon $. We can see if there is a relative increase
of this contribution, in comparison with the Born mechanism, when the
variable $Q^2$ increases.

Let us consider the asymmetries arising from the tensor polarization of the
deuteron target. The differential cross section can be written in this case
as
\be\label{Eq:eq32}
\frac{d\sigma^{(\mp )}}{d\Omega } = \frac{d\sigma_{un}^{(\mp)}}{d\Omega }
\left (1+A_{zz}^{(\mp)}(Q^2, \epsilon )Q_{zz}+A_{xz}^{(\mp)}(Q^2, \epsilon )
Q_{xz}+A_{xx}^{(\mp)}(Q^2, \epsilon )(Q_{xx}-Q_{yy})\right  ), \ee
with the following decomposition of the  asymmetries $A_{ij}^{(\mp)}(Q^2, \epsilon ), (ij=xx,xz,zz)$:
\be\label{Eq:eq33}
N^{(\mp)}(Q^2, \epsilon )A_{zz}^{(\mp)}(Q^2, \epsilon )=
A_{zz}(Q^2, \epsilon )\mp \Delta A_{zz}(Q^2, \epsilon ). 
\ee
The explicit expressions of the asymmetries as a function of FFs are:

- for the $zz$--component
\ba 
A_{zz}(Q^2, \epsilon )&=&4\tau G_Q(Q^2)[G_C(Q^2)+\frac{\tau }
{3}G_Q(Q^2)]+\frac{\tau}{2}[1+2(1+\tau )\tan^2\frac{\theta }{2}]G^2_M(Q^2),
\nn \\ \Delta A_{zz}(Q^2, \epsilon )&=&\tau [1+2(1+\tau )tan^2
\frac{\theta }{2}]G_M(Q^2)Re\Delta G_M(Q^2, \epsilon )+4\tau
[G_C(Q^2)Re\Delta G_Q(Q^2, \epsilon )+ \nn \\ &&+G_Q(Q^2)
Re(\Delta G_C(Q^2, \epsilon )+\frac{2}{3}\tau \Delta G_Q(Q^2, \epsilon ))]+
4\tau \left[(1-c^2-2c^2\tau )G_C(Q^2)+\right  .\nn \\ && \left
. +c^2\tau G_M(Q^2)-\frac{2}{3}\tau (1-c^2+4c^2\tau )G_Q(Q^2)\right ]
ReG_4(Q^2, \epsilon )+ \nn \\ && +4c\tau \sqrt{\tau (1+\tau )}
\left[G_M(Q^2)- 4(G_C(Q^2) +\frac{4}{3}\tau  G_Q(Q^2))\right]
ReG_5(Q^2, \epsilon )-\nn \\ && -2c\sqrt{\tau (1+\tau)}tan^2\frac{\theta }{2}
\left [\tau (\tau -c^2)G_M(Q^2)+\right . \nn \\ && \left .+(c^2-1)(1+2\tau )
\left (G_C(Q^2)-\frac{2}{3}\tau G_Q(Q^2)\right )\right ]
ReG_6(Q^2, \epsilon ),
\label{Eq:Eq34a} 
\ea
with
\be
\label{Eq:Eq35}
N^{(\mp)}(Q^2, \epsilon )A_{xz}^{(\mp)}(Q^2, \epsilon )=
A_{xz}(Q^2, \epsilon )\mp \Delta A_{xz}(Q^2, \epsilon ), \ee
- for the $xz$--component
\ba
A_{xz}(Q^2, \epsilon )&=&-4\tau sec\frac{\theta }{2}
\sqrt{\tau (1+\tau sin^2\frac{\theta }{2})}G_M(Q^2)G_Q(Q^2),\nn \\
\Delta A_{xz}(Q^2, \epsilon )&=&-4\tau \sec\frac{\theta }{2}
\sqrt{\tau (1+\tau \sin^2\frac{\theta }{2})}\left \{G_Q(Q^2)
Re\Delta G_M(Q^2, \epsilon )+\right . \nn \\ && +G_M(Q^2)Re\Delta
G_Q(Q^2, \epsilon )+  2[\tau (c^2-1)G_Q(Q^2)-(c^2-1+c^2\tau )G_M(Q^2)]
 \nn \\ && ReG_4(Q^2, \epsilon )-4c\sqrt{\tau(1+\tau)} \left [
G_M(Q^2)+\frac{(1-c^2)}{c^2}G_Q(Q^2)\right ] ReG_5(Q^2, \epsilon )-\nn \\
&& -\frac{1}{2c\sqrt{\tau(1+\tau)}}\left [(1-\tau +2c^2+c^2\tau )G_M(Q^2)+
\right . \nn \\ && \left .\left . +2\tau (2c^2-1)(G_M(Q^2)-G_Q(Q^2))\right ]
ReG_6(Q^2, \epsilon )\right \},
\label{Eq:eq35} \ea
with
\be\label{Eq:eq36}
N^{(\mp)}(Q^2, \epsilon )A_{xx}^{(\mp)}(Q^2, \epsilon )=
A_{xx}(Q^2)\mp \Delta A_{xx}(Q^2, \epsilon ), \ee

- and for the $xx$--component
\ba
A_{xx}(Q^2)&=&\frac{\tau }{2}G^2_M(Q^2), \nn \\ \Delta A_{xx}(Q^2, \epsilon )
&=&\tau G_M(Q^2)Re\Delta G_M(Q^2, \epsilon )+4\tau [c^2\tau G_M(Q^2)+
(c^2-1)(G_C(Q^2)- \nn \\ &&-\frac{2}{3}\tau G_Q(Q^2))]ReG_4(Q^2, \epsilon )+
4c\tau \sqrt{\tau (1+\tau )}G_M(Q^2)ReG_5(Q^2, \epsilon )+\nn \\ &&
+2c\sqrt{\frac{\tau}{(1+\tau)}} [\tau G_M(Q^2)+G_C(Q^2)-
\frac{2}{3}\tau G_Q(Q^2)]ReG_6(Q^2, \epsilon ).
\label{Eq:eq37} \ea
Note that the terms $A_{zz}(Q^2, \epsilon )$, $A_{xz}(Q^2, \epsilon )$
and $A_{xx}(Q^2)$ are the asymmetries in the Born approximation and they
coincide with the corresponding results of Ref. \cite{GM04}.

Due to the symmetry properties following from C-invariance, it is also interesting to build the sum and difference of the corresponding expressions 
for the electron and positron asymmetries due to the tensor--polarized deuteron target:
\ba
{\cal A}_{zz}^{\pm} (Q^2, \epsilon )&=&\frac{1}{2}
[A_{zz}^{(-)}(Q^2, \epsilon )\pm A_{zz}^{(+)}(Q^2, \epsilon )], \nn \\
{\cal A}_{xz}^{\pm}(Q^2, \epsilon )&=&\frac{1}{2}
[A_{xz}^{(-)}(Q^2, \epsilon )\pm A_{xz}^{(+)}(Q^2, \epsilon )], \nn \\
{\cal A}_{xx}^{\pm} (Q^2, \epsilon )&=&\frac{1}{2}
[A_{xx}^{(-)}(Q^2, \epsilon )\pm A_{xx}^{(+)}(Q^2, \epsilon )].
\label{Eq:eq37a} \ea
One can see that the quantities ${\cal A}_{ij}^+(Q^2, \epsilon )$,
$(ij=zz, xz, xx)$ do not contain the TPE contribution, neglecting
terms containing the square of the TPE amplitudes. In this approximation
we have
\ba
N{\cal A}_{zz}^+&=&A_{zz},~N{\cal A}_{xz}^+=A_{xz},
~N{\cal A}_{xx}^+=A_{xx},\nn \\ N&=&\frac{1}{2}(N^{(+)}+N^{(-)})
=A(Q^2)+B(Q^2)\tan^2\frac{\theta}{2},
\label{Eq:eq38} \ea
 coinciding with the asymmetries obtained in the Born
approximation.

The standard procedure for the determination of deuteron electromagnetic
FFs consists in measuring the unpolarized differential cross section (at
various electron scattering angles but at the same $Q^2$ value) and one
additional polarization observable (it is usually the asymmetry $A_{zz}$ due
to the tensor polarization of the deuteron target or $t_{20}$, the tensor
polarization of the recoil deuteron, with unpolarized electron beam). The
measurement of the quantity $\Sigma $, Eq. (\ref{Eq:eq30}), and of
${\cal A}_{zz}^+$ (or ${\cal A}_{xz}^+$), Eq. (\ref{Eq:eq37a}), can
be considered as the generalization of the standard procedure for extracting
electromagnetic FFs, in presence of the TPE mechanism.

On the contrary, the differences of the tensor asymmetries in the elastic
electron-- or positron--deuteron scatterings, due to the tensor polarization
of the target, are proportional to the interference of the Born amplitude and
real part of the TPE contribution. In the same approximation we have
\ba
N{\cal A}_{zz}^-&=&rA_{zz}-\Delta A_{zz}, ~N{\cal A}_{xz}^-=
rA_{xz}-\Delta A_{xz}, ~N{\cal A}_{xx}^-=rA_{xx}-\Delta A_{xx},
 \nn \\ r&=&\frac{1}{N}(\Delta A+\Delta B\tan^2\frac{\theta}{2}).
\label{Eq:eq39} \ea
The measurement of these quantities (T--even polarization observables) is
sensitive to the relative contribution of the real part of TPE
term with respect to the Born approximation.

Let us consider the double--spin asymmetries due to the longitudinal
polarization of the electron beam and the vector polarization of the
deuteron target (the transverse components of the electron spin lead to the
asymmetries suppressed by a factor $(m_e/M)$ and they are considered below).
So, the longitudinal polarization of the electron beam leads to two
asymmetries which can be written as
\be\label{Eq:eq40}
N^{(\mp)}(Q^2, \epsilon )A_{x}^{(\mp)}(Q^2, \epsilon )=
A_{x}(Q^2, \epsilon )\mp \Delta A_{x}(Q^2, \epsilon ), \ee
with
\ba
A_x(Q^2, \epsilon )&=&-2\sqrt{\tau(1+\tau)} \tan\frac{\theta }{2}
G_M(Q^2)\left [G_C(Q^2)+\frac{1}{3}\tau G_Q(Q^2)\right ], \nn \\
\Delta A_x(Q^2, \epsilon )&=&-2\sqrt{\tau(1+\tau)} \tan\frac{\theta }{2}
\left \{G_M(Q^2)Re[\Delta G_C(Q^2, \epsilon )+\frac{1}{2}(1+2\tau )
\Delta G_M(Q^2, \epsilon )+\right . \nn \\ && \left .+\frac{1}{3}\tau
\Delta G_Q(Q^2, \epsilon )\right ]-\left[G_C(Q^2)+\frac{1}{3}\tau
G_Q(Q^2)\right]Re\Delta G_M(Q^2, \epsilon )- \nn \\ && -2c^2\tau^2
G_M(Q^2)ReG_4(Q^2, \epsilon )- 4c\tau \sqrt{\tau(1+\tau)}G_M(Q^2)
ReG_5(Q^2, \epsilon )-\nn \\ && -\frac{c}{2}\sqrt{\frac{\tau}{1+\tau }}
\left [(1+\tau -c\sqrt{\tau(1+\tau)})G_M(Q^2)- \right . \nn \\
&& \left .\left .-2\left(G_C(Q^2)+\frac{1}{3}\tau G_Q(Q^2)\right )
\right]ReG_6(Q^2, \epsilon )\right \},
\label{Eq:eq40a} \ea
and
\be\label{Eq:eq41}
N^{(\mp)}(Q^2, \epsilon )A_{z}^{(\mp)}(Q^2, \epsilon )=
A_{z}(Q^2, \epsilon )\mp \Delta A_{z}(Q^2, \epsilon ), \ee
with
\ba
A_z(Q^2, \epsilon )&=&-\tau\sqrt{(1+\tau )(1+\tau sin^2\frac{\theta }{2})}
\tan\frac{\theta }{2}\sec\frac{\theta }{2}G_M^2(Q^2), \nn \\
\Delta A_z(Q^2, \epsilon )&=&-\tau\sqrt{(1+\tau )(1+\tau sin^2
\frac{\theta }{2})}\tan\frac{\theta }{2}\sec\frac{\theta }{2}G_M(Q^2)
\left \{2Re\Delta G_M(Q^2, \epsilon )+ \right .\nn \\ && +\left
[2c\sqrt{\frac{\tau}{1+\tau }}-1-c^2\right ]ReG_6(Q^2, \epsilon )+ \nn \\ &&
\left .+4\tau (1+c^2)\left [ReG_4(Q^2, \epsilon )+\frac{1}{c}
\sqrt{\frac{1+\tau }{\tau}}ReG_5(Q^2, \epsilon )\right  ]\right \}.
\label{Eq:eq41a} \ea
In the Born approximation these expressions coincide with the results of Ref. \cite{GM04} except the general sign, since in that paper  another
sign of the vector part of the deuteron spin--density matrix was taken.

Note that we can also remove or extract the TPE contribution in these
double--spin asymmetries in a similar way as it was done for the differential
cross section and the tensor asymmetries.


\section{T--odd polarization observables}

\hspace{0.7cm}


Let us consider the single--spin asymmetry induced by the transverse
polarization of the electron or positron beam. The expressions for the
spin--dependent leptonic tensor and  for the hadronic tensor, for the case
of the unpolarized final state, show that the single--spin asymmetry is
proportional to the TPE term and suppressed by the  factor ($m_e/M$).

The measurement of this small asymmetry is planned in next future \cite{Maas}. As
mentioned in the Introduction, in spite of the suppression factor, recent
measurements of the asymmetry in the scattering of transversely polarized
electrons on unpolarized protons found values different from zero, contrary
to what is expected in the Born approximation \cite{Ma05,We01}, and only one
experiment measured a single--spin observable, the recoil--deuteron vector
polarization for the elastic scattering of unpolarized electrons by
unpolarized deuteron target \cite{Pr68}.

To calculate the beam asymmetry, it is necessary to take into account also
the small amplitudes (neglected earlier, since they give a small contribution
to the other observables) which are proportional to the electron mass (the
so--called helicity flip amplitudes). The part of the matrix element of the
reaction $e^-+d\to e^-+d$, containing the helicity flip amplitudes,
can be established in analogy with the elastic nucleon--deuteron scattering
\cite{CRMS}, using the general properties of the electron--hadron interaction,
such as the Lorentz invariance and P--invariance. It can be written as
follows
\ba
{\it M}^{(flip)} &=&\frac{m_e}{M}\frac{e^2}{Q^2}\bar u(k_2)
\left [MG_{7}(s, Q^2)U_1\cdot U_2^*+\frac{1}{M}G_{8}(s, Q^2)
U_1\cdot kU_2^*\cdot k+\right  . \nn \\ && +\frac{1}{M}G_{9}(s, Q^2)
U_1\cdot pU_2^*\cdot p+\frac{1}{M}G_{10}(s,Q^2)(U_{1}\cdot kU_2^*\cdot p+
U_{2}^*\cdot kU_1\cdot p)+ \nn \\ && +\frac{1}{M}G_{11}(s,Q^2)
(U_{1}\cdot p\hat U_2^*\hat p-U_{2}^*\cdot p\hat U_1\hat p)+
\frac{1}{M}G_{12}(s,Q^2)(U_1\cdot k\hat U_2^*\hat p- \nn \\ && \left .
-U_2^*\cdot k\hat U_1\hat p-U_1\cdot kU_2^*\cdot p+U_2^*\cdot kU_1\cdot p)
\right  ]u(k_1),
\label{Eq:eq42} \ea
where all these amplitudes $G_i(s, Q^2) \ (i=7-12)$ are, in general case,
complex functions of two variables and vanish in the Born approximation
$G_i^{(Born)}(s, Q^2)=0$, $i=7-12$.

The corresponding asymmetry can be written as
\ba
&N^{(\mp)}&(Q^2, \epsilon )A_{e}^{(\mp)}(Q^2, \epsilon )=
\nn \\ && \mp \frac{4}{3}\frac{m_e}{M}tan\frac{\theta }{2}s_{y}^{(\mp)}
\left \{-\tau G_M(Q^2)\left [ImG_6(Q^2, \epsilon )+4(1+\tau )
ImG_5(Q^2, \epsilon )+\right . \right .\nn \\ && +2c\sqrt{\tau (1+\tau )}
\left (2ImG_4(Q^2, \epsilon )+ImG_{10}(Q^2, \epsilon )-\right . \nn \\
&& \left . -2(1+\tau )ImG_{11}(Q^2, \epsilon )\right )+2c^2
\left (\tau ImG_8(Q^2, \epsilon )-\right . \nn \\ && \left
.\left .-(1+\tau )
(2\tau -1)ImG_{12}(Q^2, \epsilon )\right ) \right ]+G_C(Q^2)
(ImG_6(Q^2, \epsilon )+ \nn \\ && +\frac{3}{4}ImG_7(Q^2, \epsilon ))-
\frac{2}{3}\tau G_Q(Q^2)\left [ImG_6(Q^2, \epsilon )+3\tau (1-c^2)
ImG_8(Q^2, \epsilon )\right ]+ \nn \\ &&
+\tau (G_C(Q^2)+\frac{4}{3}\tau G_Q(Q^2))\left [\frac{1}{2}
ImG_7(Q^2, \epsilon )+(1+\tau )\left (ImG_9(Q^2, \epsilon )+ \right .\right .
\nn \\ && \left . +2c^2ImG_{12}(Q^2, \epsilon )\right)+(1-c^2+c^2\tau )
ImG_8(Q^2, \epsilon )+ \nn \\ && \left .\left . +2c\sqrt{\tau (1+\tau )}
\left (ImG_{10}(Q^2, \epsilon )+\frac{1+\tau}{\tau}
Im G_{11}(Q^2, \epsilon )\right) \right ]\right \},
\label{Eq:eq43} 
\ea
where $A_e^{(-)} \ (A_e^{(+)})$ is the single--spin asymmetry (the so--called
beam asymmetry) in the scattering of transversely polarized electron
(positron) beam by unpolarized deuteron target, and 
${\vec s}^{(-)}$ $({\vec s}^{(+)})$ is the unit vector describing the polarization of the
electron (positron) beam in its rest frame. One can see that

-  $A_e^{(\pm)}$ is proportional to the electron mass and it is determined
by the electron or positron spin component perpendicular to the reaction
plane.

- $A_e^{(\pm)}$ is a T--odd observable and it vanishes in the Born
approximation as it is determined by the imaginary part of the interference
between the one-- and two--photon exchange amplitudes. Thus, the asymmetry
$A_e$ is determined by the three real electromagnetic form factors $G_M(Q^2),
G_C(Q^2), G_Q(Q^2)$ as well as by the complex TPE amplitudes:
$G_4(Q^2, \epsilon )$, $G_5(Q^2, \epsilon )$ and $G_6(Q^2, \epsilon )$
(helicity conserving) and $G_i(Q^2, \epsilon )$, $\ (i=7-12)$ (helicity non
conserving). Therefore, this observable contains all amplitudes on  equal
footing, i.e., here the helicity flip amplitudes are not suppressed in
comparison with the helicity conserving ones. Measurement of this asymmetry
in the case of elastic electron--deuteron scattering may be a more difficult
task than for the case of elastic electron--nucleon scattering due mainly 
to the fact that deuteron FFs are much smaller than nucleon FFs.

-  $A_e^{(\pm)}$ vanishes, for $\theta =0^0$ and $180^0$, as it is determined
by the product $({\vec q}\times {\vec k_1})\cdot {\vec s_e}$, and in this
case  ${\vec q}\parallel {\vec k_1}$.

Let us consider now the single--spin asymmetry due to the vector--polarized
deuteron target (the so--called target normal--spin asymmetry). The
corresponding asymmetry $A_{y}^{(\mp)}$ can be written as
\ba
&N^{(\mp)}&(Q^2, \epsilon )A_{y}^{(\mp)}(Q^2, \epsilon )=\pm\tan
\frac{\theta }{2}s_{y}\left \{ 2c\sqrt{\tau (1+\tau )}[G_M(Q^2)
Im\Delta G(Q^2, \epsilon )+  \right . 
\nn \\ 
&& +G(Q^2)Im\Delta
G_M(Q^2, \epsilon )]+2c\tau \sqrt{\tau (1+\tau )}[(1-\tau )(2c^2-1)G_M(Q^2)-
\nn \\ && 
-4(c^2-1)G(Q^2)]ImG_4(Q^2, \epsilon )-2\tau (1+\tau )[\tau (4c^2-1)
G_M(Q^2)+ 
\nn \\ && 
+4(c^2-1)G(Q^2)]ImG_5(Q^2, \epsilon )+\tau [2c
\tau \sqrt{\tau (1+\tau )}G_M(Q^2)+ \nn \\ &&  
+(4c^2-1)(G(Q^2)+G_M(Q^2))+\nn \\ &&  
+\left .2(1-\tau )((1+\tau )c^2-1)G_M(Q^2)]ImG_6(Q^2, \epsilon )\right\},
\label{Eq:eq44} \ea
where $G(Q^2, \epsilon )=G_C(Q^2, \epsilon )+\tau /3G_Q(Q^2, \epsilon ),$
${\vec s}$ is the deuteron unit polarization vector describing the vector
polarization of the target.

One can see that

- $A_{y}^{(\mp)}$ is determined by the component of the deuteron polarization
vector perpendicular to the reaction plane, i.e., by the following product
$({\vec k}_1\times {\vec k}_2)\cdot {\vec s}.$

- $A_{y}^{(\mp)}$ vanishes when ${\vec k}_2|| {\vec k}_1,$ i.e., for forward
and backward scatterings.

- $A_{y}^{(\mp)}$ is a T--odd observable and it is zero in the Born
approximation. It is determined by the interference of the one-- and
two--photon exchange amplitudes (by the imaginary parts of all six complex
TPE  helicity conserving  amplitudes).

At last let us consider the scattering of the longitudinally polarized
electron or positron beam by the tensor--polarized deuteron target. In this
case, the asymmetries can be written as follows:
\ba
N^{(\mp)}(Q^2, \epsilon )A_{xy}^{(\mp)}(Q^2, \epsilon )&=&\mp
2\sqrt{\frac{\tau }{1+\tau}}\left \{ \left [\tau G_M(Q^2)-2G_C(Q^2)+
\frac{4}{3}\tau G_Q(Q^2)\right ]ImG_6(Q^2, \epsilon )+ \right .  \nn
\\ &&\left . +4\tau G_M(Q^2)Im \left [c\sqrt{\tau (1+\tau )}G_4(Q^2, \epsilon )
+(1+\tau )G_5(Q^2, \epsilon )\right ]\right \}, \nn
\\ N^{(\mp)}(Q^2, \epsilon )A_{yz}^{(\mp)}(Q^2, \epsilon )&
=&\mp 2\tau \sqrt{\frac{\tau }{1+\tau}}\tan\frac{\theta }{2}
\left \{2(1+\tau )\left[G_Q(Q^2)Im\Delta G_M(Q^2, \epsilon )-
\right . \right .\nn \\ &&
\left . -G_M(Q^2)Im\Delta G_Q(Q^2, \epsilon )\right ]+
\nn \\ && +4c(1+\tau )\sqrt{\tau }G_M(Q^2) Im \left[c\sqrt{\tau}
G_4(Q^2, \epsilon )+2\sqrt{(1+\tau )}G_5(Q^2, \epsilon )\right ]+
\nn  \\ && \left . +c\sqrt{\frac{1+\tau }{\tau }}\left [2\tau G_Q(Q^2)+
(2\tau -1)G_M(Q^2)\right ]ImG_6(Q^2, \epsilon )\right \}.
\label{Eq:eq45} \ea
Note that the asymmetry $A_{xy}$ is determined by the imaginary parts of the amplitudes $G_{4,5,6}$,  which differ in spin structure from the Born spin structure. Both asymmetries
are zero in the Born approximation since they are determined by the
interference of the one-- and two--photon exchange mechanisms.

The polarized deuteron targets, generally used in high--energy experiments,
have zero $Q_{xy}$ and $Q_{yz}$ parameters, since the polarization state is
determined by the population numbers $n_{\pm ,0}$, i.e., by the diagonal
elements of the spin--density matrix  of the deuteron target. The
determination of these asymmetries requires a polarized deuteron targets
with non--zero $Q_{xy}$ and $Q_{yz}$ parameters or the measurement of the
corresponding components of the tensor polarization of the recoil deuteron
(the target in this case is unpolarized).

In polarization experiments it is possible to prepare the deuteron target
with polarization along (opposite) a definite direction. In our case the
natural direction is the virtual photon momentum (or $z$ axis). Similar
polarization effects were considered in Ref. \cite{GP64}: longitudinal and
transverse polarizations of the recoil deuteron in the elastic
electron--deuteron scattering.

Let us consider the case when the spin of the deuteron target has definite
projection on $z$ axis. It is convenient in this case to write the
contraction of the leptonic and hadronic tensors in the following general
form
\be\label{Eq:eq1a}
S=L_{\mu\nu}H_{\mu\nu}=S_{\mu\nu}U_{\mu}U_{\nu}^*, \ee
where $U_{\mu}$ is the polarization four--vector of the deuteron target.
Then, with unpolarized electron beam, the $S_{\mu\nu}$ tensor can be written
as
\be\label{Eq:eq2a}
S_{\mu\nu}=S_1g_{\mu\nu}+S_2q_{\mu}q_{\nu}+S_3k_{\mu}k_{\nu}+
S_4(q_{\mu}k_{\nu}+q_{\nu}k_{\mu})+iS_5(q_{\mu}k_{\nu}-q_{\nu}k_{\mu}),
\ee
where the structure functions $S_i$, $(i=1-5)$, can be expressed in terms of
the generalized form factors $G_i(Q^2, \epsilon )$, $i=M, C, Q,$ and of the
amplitudes $G_i(Q^2, \epsilon )$, $i=4, 5, 6,$ as follows:
\ba
S_1&=&-q^4\left [(1+\tau )|G_M|^2+\frac{1}{\tau}\cot^2\frac{\theta}{2}
|G_C-\frac{2}{3}\tau G_Q|^2+2c\sqrt{\tau (1+\tau )}ReG_MG_6^*\right],
\nn \\ S_2&=&-q^2\left(1+\tau +\cot^2\frac{\theta}{2}\right)\left [|G_M|^2
+4\frac{\tau}{1+\tau}ReG_MG_Q^*\right]-4\frac{q^2}{1+\tau}\cot^2
\frac{\theta}{2}\left [\frac{\tau}{3}|G_Q|^2+  \right .\nn \\ && \left .
+ReG_CG_Q^*\right ]+4cq^2\sqrt{\frac{\tau}{1+\tau}}Re\left\{G_M\left
[2(1+\tau )(1-c^2+\tau c^2)G_5+G_6\right ]^*+ \right . \nn \\ && \left .
+G_C\left [4\cot^2\frac{\theta}{2}G_5-(2+\tau )G_6\right ]^*+\frac{\tau}{3}
G_Q\left [4\cot^2\frac{\theta}{2}G_5+(1+2\tau )G_6\right ]^*\right \},
\nn \\ S_3&=&4M^2\tau (1+\tau )ReG_M\left [G_M+8(\tau G_4+c\sqrt{\tau
(1+\tau )}G_5)\right ]^*+ \nn \\ && +16M^2Re(G_C-\frac{2}{3}\tau G_Q+
\tau G_M)\left [2\tau\cot^2\frac{\theta}{2}G_4+c\sqrt{\tau (1+\tau )}G_6
\right ]^*,  \nn \\ S_4&=&cq^2\sqrt{\tau (1+\tau )}\left [|G_M|^2+
2ReG_MG_Q^*\right ]+  \nn \\ && +q^2Re\left \{ G_M\left [-4c\sqrt{\tau (1+
\tau )} \left (3\tau +\frac{1-\tau}{1+\tau}\cot^2\frac{\theta}{2}\right )G_4
+8\tau \left(1+\tau +\cot^2\frac{\theta}{2}\right )G_5+ \right .\right .
\nn \\ &&\left. +\left(2\tau -3-2\frac{1-\tau}{1+\tau}\cot^2\frac{\theta}{2}
\right )G_6\right ]^*+2\left (G_C-\frac{2}{3}\tau G_Q \right )\left [4c
\sqrt{\frac{\tau}{1+\tau }}\cot^2\frac{\theta}{2}G_4+ \right . \nn
\\ && \left . +\left(2-\tau +\frac{2}{1+\tau}\cot^2\frac{\theta}{2}\right)G_6
\right ]^*+2\tau G_Q\left [4c\sqrt{\frac{\tau}{1+\tau }}\cot^2
\frac{\theta}{2}G_4+ \right . \nn \\ && \left .\left . +4\cot^2
\frac{\theta}{2}G_5+\left (1+\frac{2}{1+\tau}\cot^2\frac{\theta}{2}\right )
G_6\right ]^* \right \},  \nn \\ S_5&=&8cM^2\sqrt{\tau (1+\tau )}
ImG_M(G_C+\frac{\tau}{3}G_Q)^*+  \nn \\ && +q^2Im\left  \{4\cot^2
\frac{\theta}{2}\left(G_C+\frac{\tau}{3}G_Q\right) \left [2G_5+
\frac{1}{1+\tau}\left(2c\sqrt{\tau (1+\tau )}G_4+G_6 \right)\right ]^*+
\right . \nn \\ && +2\tau \left [(1+\frac{2}{3}\tau )G_Q-G_C\right ]G_6^*+
G_M\left [-12c\tau\sqrt{\tau(1+\tau)}G_4+(2\tau -3)G_6- \right .\nn
\\ && \left . \left . -2\frac{1-\tau}{1+\tau}\cot^2\frac{\theta}{2}
(2c\sqrt{\tau(1+\tau)}G_4+G_6)\right ]^*\right\}.  \label{Eq:eq48} \ea
The contraction of the leptonic and hadronic tensors in the case when
the polarization of the deuteron target has definite projection on the $z$ axis, $m=\pm 1, 0$, can be written as
\be\label{Eq:eq49}
S^{(m)}=S_{\mu\nu}U_{\mu}^{(m)}U_{\nu}^{(m)*}, \ee
where $U_{\mu}^{(m)}$ is the deuteron polarization four--vector with
projection $m$
$$U_{\mu}^{(\pm)}=\mp \frac{1}{\sqrt{2}}(0, 1, \pm i, 0),
\ U_{\mu}^{(0)}=(0, 0, 0, 1). $$
After straightforward calculations we obtain
\be\label{Eq:eq50}
S^{(+)}=S^{(-)}=-S_1+\frac{1}{2}\frac{Q^2}{1+\tau}\cot^2\frac{\theta}{2}S_3,
\ee
$$S^{(0)}=-S_1+(1+\tau )Q^2S_2+Q^2(1+\tau +\cot^2\frac{\theta}{2})
\frac{\tau}{1+\tau}S_3+2c\sqrt{\tau (1+\tau )}Q^2S_4. $$
It follows that the cross sections of the elastic electron--deuteron scattering can be
written in the familiar form
\be\label{Eq:eq51}
\frac{d\sigma^{(m)}}{d\Omega}=\sigma_0\left [A^{(m)}+
B^{(m)}\tan^2\frac{\theta}{2}\right  ]. \ee
Let us separate the dominant (Born) and TPE contributions to these
structure functions and define
\be\label{Eq:eq52}
A^{(m)}=A_B^{(m)}+\Delta A^{(m)}, \ \
B^{(m)}=B_B^{(m)}+\Delta B^{(m)}, \ee
where the index $B$ indicates the Born contribution. The Born terms can be
expressed in terms of the deuteron FFs as:
\ba
A_B^{(+)}&=&A_B^{(-)}=\frac{\tau}{2}G_M^2(Q^2)+\left [G_C(Q^2)-
\frac{2\tau}{3}G_Q(Q^2)\right  ]^2,~\nn
\\ B_B^{(+)}&=&B_B^{(-)}=\tau (1+\tau )G_M^2(Q^2),\nn
\\ A_B^{(0)}&=&\tau G_M^2(Q^2)+G_C^2(Q^2)+\frac{8}{3}\tau G_C(Q^2)G_Q(Q^2)+
\frac{16}{9}\tau^2G_Q^2(Q^2), \nn  \\ B_B^{(0)}&=&2\tau (1+\tau )
G_M^2(Q^2). \label{Eq:eq53} \ea
Summing the structure functions over the index $m$ (i.e., over all possible
deuteron spin projections) and dividing by three (the averaging over the
deuteron spin) we obtain the usual structure functions $A(Q^2)$ and $B(Q^2)$.

Gourdin and Piketty calculated the longitudinal and two transverse
polarizations of the recoil deuteron in the elastic electron--deuteron
scattering \cite{GP64}. The two transverse polarizations are orthogonal to
the recoil deuteron momentum (it is ${\vec q}$ in lab. system), one in the
scattering plane (along $x$ axis in our case) and the other one normal to
this plane (along $y$ axis). The quantities $S^{(x,y)}$, corresponding to
the deuteron target polarized along $x$ and $y$ directions, are
\be\label{Eq:eq54}
S^{(x)}=-S_1+\frac{Q^2}{1+\tau}\cot^2\frac{\theta}{2}S_3,
\ \ S^{(y)}=-S_1, \ee
and the corresponding structure functions $A^{(x,y)}$ and $B^{(x,y)}$ can
be written as (in the Born approximation)
\ba
A_B^{(x)}&=&\tau G_M^2(Q^2)+\left [G_C(Q^2)-\frac{2\tau}{3}
G_Q(Q^2)\right  ]^2, \ B_B^{(x)}=\tau (1+\tau )G_M^2(Q^2), \nn
\\ A_B^{(y)}&=&(G_C(Q^2)-\frac{2\tau}{3}G_Q(Q^2))^2,
\ B_B^{(y)}=\tau (1+\tau )G_M^2(Q^2). \label{Eq:eq55}\ea
This result for $A_B^{(i)} $, $B_B^{(i)} $, $i=x,y$ coincides with the
results obtained in Ref. \cite{GP64}. Following Ref. \cite{GP64}, it is
convenient to introduce a total transverse differential cross section
$\sigma_T$ and calculate the difference $D_T=(\sigma_T^{(+)}+
\sigma_T^{(-)}-\sigma_T^{(0)})$. An interesting result is the independence
of  $D_T$ from the  the deuteron magnetic FF $G_M$ and consequently with
respect to the scattering angle
\be\label{Eq:eq56}
\frac{1}{\sigma_0}(\sigma^{(+)}+\sigma^{(-)}-\sigma^{(0)})=
G_C^2(Q^2)-\frac{16}{3}\tau G_C(Q^2)G_Q(Q^2)-\frac{8}{9}\tau^2 G_Q^2(Q^2).
\ee
This expression (valid in the Born approximation) coincides also with the
result of Ref. \cite{GP64}.

The explicit expressions of the corrections $\Delta A^{(m)}$ and
$\Delta B^{(m)}$ due to the TPE contributions are
\ba
\Delta A^{(+)}&=&\Delta A^{(-)}=\mp \tau G_M(Q^2)Re \left
\{\Delta G_M(Q^2, \epsilon)+4\tau c^2G_4(Q^2, \epsilon)+ \right . \nn
\\ 
&& 
\left .+2c\sqrt{\tau (1+\tau )}\left [2G_5(Q^2, \epsilon)+
\frac{1}{1+\tau}G_6(Q^2, \epsilon) \right ]\right  \}\mp 2\left [G_C(Q^2)-
\frac{2}{3}\tau G_Q(Q^2)\right ] \nn  \\ && Re\left [\Delta G_C(Q^2, \epsilon)
-\frac{2}{3}\tau\Delta G_Q(Q^2, \epsilon)+2\tau (c^2-1)G_4(Q^2, \epsilon)+
c\sqrt{\frac{\tau}{1+\tau}}G_6(Q^2, \epsilon)\right ], \nn  \\
\Delta B^{(+)}&=&\Delta B^{(-)}=\mp 2\tau \sqrt{1+\tau}G_M(Q^2)Re\left
[\sqrt{1+\tau}\Delta G_M(Q^2, \epsilon)+c\sqrt{\tau}G_6(Q^2, \epsilon)
\right ], \nn  \\ \Delta A^{(0)}&=&\mp \left\{ 2\tau G_M(Q^2)Re
\Delta G_M(Q^2, \epsilon)+2G_C(Q^2)Re\left [\Delta G_C(Q^2, \epsilon)+
\frac{4}{3}\tau \Delta G_Q(Q^2, \epsilon)\right ]+\right . \nn  \\ 
&&
\left .
+\frac{8}{3}\tau G_Q(Q^2)Re\left [\Delta G_C(Q^2, \epsilon)+\frac{2}{3}\tau
\Delta G_Q(Q^2, \epsilon)\right ] \right\}\mp  \nn  \\ && \mp 4\tau G_M(Q^2)
Re\left\{ 2\tau \left(1+4\tau +\frac{2}{1+\tau}\cot^2\frac{\theta}{2}
\right)G_4(Q^2, \epsilon)+\right . \nn  \\ 
&& 
\left . +c\sqrt{\tau (1+\tau )}
\left [2(1+2\tau )G_5(Q^2, \epsilon)+  \frac{1}{1+\tau}G_6(Q^2, \epsilon)
\right ]\right\}\pm \nn  \\ 
&&
\pm 4\tau \left [G_C(Q^2)+\frac{4}{3}\tau G_Q(Q^2)
\right ]Re\left\{ 2\tau \left(1+\frac{1}{1+\tau}\cot^2\frac{\theta}{2}
\right )G_4(Q^2, \epsilon)+ \right  .\nn  \\ && \left .+c\sqrt{\tau (1+\tau )}
\left [4G_5(Q^2, \epsilon)+\frac{1}{1+\tau}G_6(Q^2, \epsilon)\right ]
\right\},  \nn  
\\ \Delta B^{(0)}&=&\mp 4\tau (1+\tau )G_M(Q^2)Re\left [
\Delta G_M(Q^2, \epsilon)-\right .
\nn  \\ 
&& \left .-2c(1+\tau )\sqrt{\tau (1+\tau )}G_5(Q^2, \epsilon)+
8\tau^2G_4(Q^2, \epsilon) \right ]\mp  \nn  \\ 
&&  
\mp 4\tau c\sqrt{\tau (1+\tau )}
\left [G_M(Q^2)+(1+2\tau )(G_C(Q^2)-\frac{2}{3}\tau G_Q(Q^2))\right ]
ReG_6(Q^2, \epsilon), \label{Eq:eq57} \ea
where the signs $(\mp )$ indicate the scattering of the electron (positron)
by a polarized deuteron target.

\section{Conclusion}

\hspace{0.7cm}

Precise measurements of various observables in the elastic electron--proton
scattering arose the question of the importance of the TPE mechanism. In
this work, the study of TPE contribution and its consequences on the
extraction of hadron electromagnetic FFs has been extended to
electron--deuteron elastic scattering. The determination of the deuteron
electromagnetic FFs from the measurement of the differential cross section
and one polarization observable in the elastic electron--deuteron scattering
is valid only in Born approximation. In case of deuteron and light nuclei
the relative contribution of TPE term with respect to the main term, the
$1\gamma$ exchange, is expected to be larger at the same momentum transfer squared, due to the steeper decrease of the FFs. Therefore, it can be detectable at smaller  $Q^2$ values than in the case of
the elastic electron--proton scattering.

A model--independent analysis of the influence of the two--photon exchange
mechanism on the differential cross section and on various polarization
observables has been performed for the elastic electron (positron)--deuteron
scattering. General symmetry properties of the electromagnetic
lepton--hadron interaction (as the lepton helicity conservation in QED at
high energies, the C--invariance and crossing symmetry) were used in this
analysis. These properties allows to parametrize the amplitudes of
$e^{\mp}d-$scattering  in terms of fifteen real functions, in presence of
the TPE mechanism: three standard electromagnetic deuteron FFs, which are
the functions of one variable $Q^2$, and six complex functions that depend
on two variables, $Q^2$ and $\epsilon $. The expressions for
the differential cross section and all polarization observables have been given in terms of
these functions. We have considered the case of an arbitrary polarized
deuteron target and polarized electron beam (both longitudinal and transverse).
The transverse polarization of the electron beam leads to a single--spin
asymmetry which is non--zero in presence of the two--photon exchange
contribution but it is suppressed by the factor ($m_e/M$). Let us note that
this factor is appreciably larger in the case of muon--deuteron scattering.

It was shown that the measurements of the differential cross section and
one polarization observable (for example, the tensor asymmetry) for electron
and positron deuteron elastic scattering, in the same kinematical conditions,
allows to extract the deuteron electromagnetic form factors. 

All the results derived in this work hold when the terms proportional to the square of the
two--photon exchange amplitudes are neglected.

\section{Acknowledgment}

This work was inspired by enlightening discussions with Prof. M. P. Rekalo.
One of us (G.I.G.) acknowledges the hospitality of CEA, Saclay, where part of this work was done.


\begin{thebibliography}{99}
\bibitem{Re68}
  A.~I.~Akhiezer and M.~P.~Rekalo,
  Sov.\ Phys.\ Dokl.\  {\bf 13} (1968) 572
  [Dokl.\ Akad.\ Nauk Ser.\ Fiz.\  {\bf 180} (1968) 1081]; 
  A.~I.~Akhiezer and M.~P.~Rekalo,
  Sov.\ J.\ Part.\ Nucl.\  {\bf 4} (1974) 277
  [Fiz.\ Elem.\ Chast.\ Atom.\ Yadra {\bf 4} (1973) 662]
\bibitem{Jo00}
M. K. Jones {\it et al.}, Phys. Rev. Lett. 84 (2000) 1398;
O. Gayou {\it et al.}, Phys. Rev. Lett. 88 (2002) 092301;
  V.~Punjabi {\it et al.},
  Phys.\ Rev.\ C 71 (2005) 055202.
 [Erratum-ibid.\ C {\bf 71} (2005) 069902].
\bibitem{AC}
L.~Andivahis {\it et al.},
Phys.\ Rev.\ D {\bf 50}, 5491 (1994);
M.~E.~Christy {\it et al.}  [E94110 Collaboration],
Phys.\ Rev.\ C {\bf 70}, 015206 (2004);
R.G. Arnold {\it et al.}, Phys. Rev. Lett. {\bf 35}, 776 (1975).
\bibitem{Qa05}
I.~A.~Qattan {\it et al.},
Phys.\ Rev.\ Lett.\  {\bf 94}, 142301 (2005).

\bibitem{Gu03}
P.~A.~M.~Guichon and M.~Vanderhaeghen,
Phys.\ Rev.\ Lett.\  {\bf 91}, 142303 (2003).
\bibitem{Bl03}
  P.~G.~Blunden, W.~Melnitchouk and J.~A.~Tjon,
  Phys.\ Rev.\ C {\bf 72}, 034612 (2005)

\bibitem{Ch04}
  A.~V.~Afanasev, S.~J.~Brodsky, C.~E.~Carlson, Y.~C.~Chen and M.~Vanderhaeghen,
  Phys.\ Rev.\ D {\bf 72}, 013008 (2005)

\bibitem{PCV}
  V.~Pascalutsa, C.~E.~Carlson and M.~Vanderhaeghen,
  Phys.\ Rev.\ Lett.\  {\bf 96}, 012301 (2006)
\bibitem{Ko05}
  S.~Kondratyuk, P.~G.~Blunden, W.~Melnitchouk and J.~A.~Tjon,
  Phys.\ Rev.\ Lett.\  {\bf 95}, 172503 (2005)
\bibitem{By06}
  Yu.~M.~Bystritskiy, E.~A.~Kuraev and E.~Tomasi-Gustafsson,
  arXiv:hep-ph/0603132; 
\bibitem{Ku06}
  E.~A.~Kuraev, V.~V.~Bytev, Yu.~M.~Bystritskiy and E.~Tomasi-Gustafsson,
  Phys.\ Rev.\ D {\bf 74} (2006) 013003.
\bibitem{Bo06}
  D.~Borisyuk and A.~Kobushkin,
  arXiv:nucl-th/0606030.
\bibitem{TG05}
  E.~Tomasi-Gustafsson and G.~I.~Gakh,
  Phys.\ Rev.\ C {\bf 72}, 015209 (2005).



\bibitem{Re04a}
M.~P.~Rekalo and E.~Tomasi-Gustafsson,
Eur.  Phys. J. A. {\bf 22}, 331 (2004).
\bibitem{Re04b}
M.~P.~Rekalo and E.~Tomasi-Gustafsson,
Nucl.\ Phys.\ A {\bf 740}, 271 (2004).
\bibitem{Re04c}
M.~P.~Rekalo and E.~Tomasi-Gustafsson,
Nucl.\ Phys.\ A {\bf 742}, 322 (2004).
\bibitem{Ga05}
  G.~I.~Gakh and E.~Tomasi-Gustafsson,
  Nucl.\ Phys.\ A {\bf 761}, 120 (2005).

\bibitem{Ga06}
  G.~I.~Gakh and E.~Tomasi-Gustafsson,
  Nucl.\ Phys.\ A {\bf 771}, 169 (2006).
\bibitem{Gu73}
J. Gunion, L. Stodolsky, Phys. Rev. Lett. {\bf 30}, 345
(1973).
\bibitem{Fr73}
V. Franco, Phys. Rev. {\bf 8}, 826 (1973).
\bibitem{Bo73}
V. N. Boitsov, L.A. Kondratyuk, and V. B. Kopeliovich,
Sov. J. Nucl. Phys. {\bf 16}, 287 (1973);
F. M. Lev, Sov. J. Nucl. Phys. {\bf 21}, 45 (1973).

\bibitem{Re99}
M. P. Rekalo, E. Tomasi-Gustafsson and D. Prout, Phys. Rev.
{\bf  C60}, 042202 (1999).
\bibitem{Al99} L. C. Alexa {\it et al.}, Phys. Rev.  Lett. {\bf 82}, 1374 
(1999).
\bibitem{Ab99} D. Abbott {\it et al.}, Phys. Rev. Lett {\bf 82}, 1379 (1999).

\bibitem{H98}
A. Huber {\it et al.}, Phys. Rev. Lett. {\bf 80}, 468(1998).

\bibitem{We01}
S. P. Wells et al. [SAMPE collaboration], Phys. Rev. C {\bf 63}, 064001 (2001).

\bibitem{Ma05} 
  F.~E.~Maas {\it et al.},
  Phys.\ Rev.\ Lett.\  {\bf 94}, 082001 (2005).


\bibitem{Pr68}
R. Prepost, R. M. Simonds and B. H. Wiik, Phys. Rev. Lett. {\bf 21}, 1271 (1968).

\bibitem{CRMS}
J. N. Chahoud, G. Russo, P. Mazzanti and M. Salvini,
Nuovo Cim. A  {\bf 56}, 838 (1968).
\bibitem{MT}
P. J. Mohr and B. N. Taylor, Rev. Mod. Phys. {\bf 72}, 351 (2000).
\bibitem{ERC}
T. E. O. Ericson and M. Rosa--Clot, Nucl. Phys. A {\bf 405}, 497 (1983).


\bibitem{Ga06a}
  G.~I.~Gakh, E.~Tomasi-Gustafsson, C.~Adamuscin, S.~Dubnicka and A.~Z.~Dubnickova,
  Phys.\ Rev.\ C {\bf 74}, 025202 (2006).


\bibitem{GG02}
  R.~A.~Gilman and F.~Gross,
  J.\ Phys.\ G {\bf 28}, R37 (2002)
\bibitem{GM04}
  G.~I.~Gakh and N.~P.~Merenkov,
  J.\ Exp.\ Theor.\ Phys.\   {\bf 98}, 853(2004).
  [Zh.\ Eksp.\ Teor.\ Fiz.\  {\bf 98}, 982 (2004)].


\bibitem{Maas} F. Maas, private communication.
\bibitem{GP64}
M. Gourdin and C. A. Piketty, Nuovo Cim.  {\bf 32}, 1137 (1964).
\end{thebibliography}
\end{document}